# Atomically smooth single-crystalline platform for low-loss plasmonic nanocavities

Lufang Liu[1,#], Alexey V. Krasavin[2,#], Junsheng Zheng[1], Yuanbiao Tong[1], Pan Wang[1,*], Xiaofei Wu[3], Bert Hecht[3], Chenxinyu Pan[1], Jialin Li[1], Linjun Li[1], Xin Guo[1], Anatoly V. Zayats[2,*], Limin Tong[1,*]

[1]State Key Laboratory of Modern Optical Instrumentation, College of Optical Science and Engineering, Zhejiang University, Hangzhou 310027, China

[2]Department of Physics and London Centre for Nanotechnology, King's College London, Strand, London WC2R 2LS, UK

[3]NanoOptics & Biophotonics Group, Experimentelle Physik 5, Physikalisches Institut, Universität Würzburg, Am Hubland, 97074 Würzburg, Germany

[#]These authors contributed equally to this work

*Corresponding authors. Email: nanopan@zju.edu.cn, a.zayats@kcl.ac.uk, phytong@zju.edu.cn

**ABSTRACT:** Nanoparticle-on-mirror plasmonic nanocavities, capable of extreme optical confinement and enhancement, have triggered state-of-the-art progress in nanophotonics and development of applications in enhanced spectroscopies and molecular detection. However, the optical quality factor and thus performance of these nanoconstructs are undermined by the granular polycrystalline metal films used as a mirror. Here, we report an atomically smooth single-crystalline platform for low-loss nanocavities using chemically-synthesized gold microflakes as a mirror. Nanocavities constructed using gold nanorods on such microflakes exhibit a rich structure of plasmonic modes, which are highly sensitive to the thickness of optically-thin (down to ~15 nm) microflakes. The atomically smooth single-crystalline microflakes endow nanocavities with significantly improved quality factor (~2 times) and scattering intensity (~3 times) compared with their counterparts based on deposited films. The developed low-loss nanocavities further allow for the integration with a mature platform of fiber optics, opening opportunities for realizing nanocavity-based miniaturized photonic devices with high performance.

The ability to truly confine and modulate light on the nanoscale allows access to the regime of extreme light-matter interactions for fundamental studies as well as the realization of highly compact nanophotonic devices. By coupling optical fields with collective electronic excitations (i.e., surface plasmons), plasmonic nanoparticles have the ability to confine light down to deep-subwavelength scale (e.g., 10 nm) and produce enhanced local electromagnetic fields[1,2]. However, it is challenging to achieve more tightly confined optical fields (e.g., sub-5 nm)[3]. Recently, nanoparticle-on-mirror (NPoM) plasmonic nanocavity[4,5], formed by placing a metal nanoparticle on a metal film separated with a nanometer-thick dielectric layer, has attracted intensive research interests due to its capability of extreme optical confinement[6-9], facile incorporation of functional materials into the gap and ease of fabrication. They have given rise to a series of breakthrough in state-of-the-art nanophotonic research and applications[10-29], such as spontaneous emission enhancement[13,15,17], strong coupling[16,18,25], optical sensing[11,19,27,28], and quantum plasmonics[20,21]. Usually, the implementation of NPoM nanocavities uses deposited metal films as the mirror[6-9,11-19,21-29], which have a polycrystalline structure and a typical surface root-mean-square (RMS) roughness of a few nanometers[4]. Due to the extreme confinement of optical fields in the nanometer-scale gap, granular polycrystalline metal films can introduce a significant optical loss because of the scattering of electrons by surface roughness and numerous grain boundaries[30-33]. This limits the optical quality of NPoM nanocavities as well as prevents from achieving and exploiting multiresonant nanocavities as closely spaced resonances become merged due to the low quality factors. The poor surface quality can further cause a deviation of the optical response of an NPoM nanocavity from the designed parameters and significant cavity-to-cavity variation of the optical response due to the fluctuation in the gap geometry and thickness[34], which inevitably degrades the performance of nanocavities. Therefore, improving the structural quality of metal mirrors is critical to exploit the full potential of NPoM nanocavities. By using template-stripped metal films[35], the RMS roughness of the metal surface in contact with the ultrasmooth template can be as low as ~0.2 nm[36], which is beneficial for the improvement of optical quality of NPoM nanocavities[4,15,22]. However, as-deposited metal films retain the polycrystalline structure, affecting optical performance[33,37]. The rough surface at the opposite side of the film still has an obvious effect on the optical quality of nanocavities constructed on optically-thin metal films, especially when the access through the film is needed, as in the case of excitation with total internal reflection approach[38,39] or in integrated optics. In addition, with this approach it is challenging to fabricate an integrated metallic mirror in a target device, which is highly required for the miniaturization and integration of NPoM nanocavity-based devices.

Here, by using atomically smooth single-crystalline gold microflakes (GMFs, with tunable thickness down to ~15 nm) as a mirror, we report a new platform for NPoM nanocavities with superior optical properties in terms of GMF thickness-dependent mode structure tunability and significantly improved quality factor (~2 times) and scattering intensity (~3 times) compared with their counterparts based on deposited films. The use of plasmonic nanorods to define a cavity further provides multiresonant response, in contrast to nanospheres, important for nonlinear optical and sensing applications. Moreover, the transferable GMFs allow for the facile integration of low-loss nanocavities with mature platform of fiber optics, where optically-thin ultrasmooth mirrors allow excitation through the substrate, opening new opportunities for developing nanocavity-based photonic devices with miniaturized sizes.

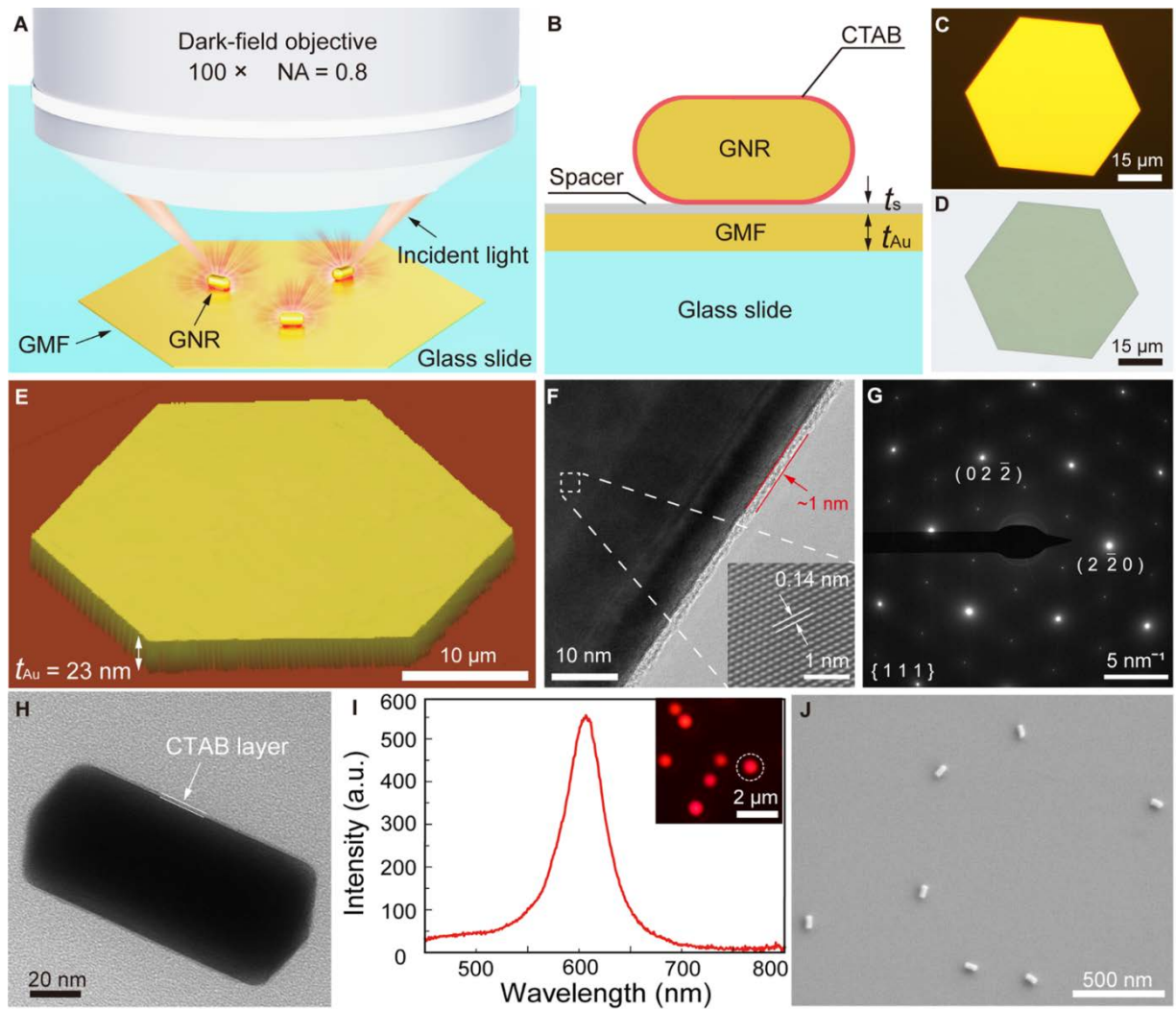

**Figure 1.** Design of NRoMF nanocavities. **A,** Schematic representation and **B,** a cross-sectional view of a GNR on a GMF separated by a dielectric spacer. Illumination condition in a dark-field microscope is also shown in **A**. **C,D,** Optical micrographs of a GMF ($t_{Au}$ = 27 nm) taken in (**C**) reflection and (**D**) transmission. **E,** AFM image of a GMF. **F,** TEM image of an edge of a GMF showing the existence of a native organic layer. Inset: high-resolution TEM image of the GMF. **G**, Corresponding electron diffraction pattern of the GMF. **H,** TEM image of a GNR. **I,** Scattering spectrum of a GNR on a glass slide under unpolarized excitation and detection. Inset: dark-field scattering image of the measured GNR (circled). **J,** Scanning electron microscopy (SEM) image of NRoMF nanocavities formed on a GMF.

**RESULTS AND DISCUSSION**

**Fabrication of nanocavities.** In order to achieve multiresonant high-Q nanocavities capable of operating in a broad spectral range[40], gold nanorods (GNRs) were used to define cavities on GMFs, i.e., nanorod-on-microflake (NRoMF) nanocavities (Fig. 1**A**,**B**). Single-crystalline GMFs, with thicknesses tunable in the range of ~10-300 nm, were chemically-grown on a glass slide[41,42] (Supporting Information Section 1). Figure 1**C** shows an optical micrograph of a GMF with a thickness of 27 nm, which is semitransparent and gives a greenish color in transmission due to the relatively small thickness (Fig. 1**D**). Figure 1**E** presents an atomic force microscopy (AFM) image of a GMF with a lateral size of ~37 μm, which indicates that as-synthesized GMF has an atomically smooth surface (RMS roughness ~0.2 nm) and a constant thickness (~23 nm) across the whole flake. A transmission electron microscopy (TEM) image of an edge of a GMF shows the existence of a native organic layer (~1 nm in thickness) on the surface of GMFs (Fig. 1**F**). Clear lattice fringes with the atomic planes spaced by 0.14 nm can be observed (Fig. 1**F**, inset). Together with the electron diffraction pattern (Fig. 1**G**), it confirms that the GMF is single crystalline and has a preferential growth direction along the [111] axis[42]. The excellent surface quality of GMFs together with the single crystallinity and tunable thickness makes GMFs extremely attractive for high-performance nanocavities.

GNRs, with average length and diameter of 105 and 55 nm, were synthesized by a seed-mediated method (Supporting Information Section 2). Each GNRs is capped with a bilayer of cetyltrimethylammonium bromide (CTAB) having a thickness of ~1 nm (Fig. 1**H**), which can prevent a direct contact between GNRs and metallic mirrors (Fig. 1**B**). When deposited on a glass slide, individual GNRs appear as red spots in a dark-field scattering image (Fig. 1**I**, inset), showing the longitudinal surface plasmon resonance peak blue-shifted to around 610 nm (cf., Fig. 1**I** and Fig. S2**B**). By drop-casting a diluted solution of GNRs onto GMFs on a glass slide, well-separated individual NRoMF nanocavities, with a total gap distance of ~2 nm, can be obtained (Fig. 1**J**).

**Cavity mode identification.** Different from the scattering image of GNRs on a glass substrate, the nanocavities formed on a 100-nm-thickness GMF (the thickness much larger than the skin depth of gold) under the oblique illumination with an unpolarized white light (Fig. 1**A**, see Supporting Information Section 3 for details) exhibit a distinct scattering image consisting of both green (in the middle) and red (with a doughnut-shaped spatial distribution) scattering components in each nanocavity (cf. inset of Fig. 1**I** and Fig. 2**A**,**B**). Correspondingly, the measured scattering spectrum (Fig. 2**C**, black line) reveals three scattering peaks, locating at wavelengths of ~550, 651 and 709 nm, respectively.

To identify the observed modes, finite element numerical simulations of near-field scattering of the NRoMF nanocavities (Supporting Information Section 4) were performed for plane wave illumination at different nanorod orientation and light polarization configurations (inset of Fig. 2**D**). This allows reconstruction of the experimental configuration with randomly oriented nanorods in the sample plane under unpolarized incident light by averaging plane wave scattering signals. Under $TM_1$ excitation (Fig. 2**D**, blue line), there are four resonance peaks located at wavelengths of 565, 660, 730 and 928 nm, labeled as modes $M_2$, $V_1$, $V_2$ and $M_4$, respectively. However, in the case of $TE_1$ and $TE_2$ excitations (Fig. 2**D**, violet and yellow lines), modes $V_1$ and $V_2$ are absent. This indicates that they correspond to vertically polarized resonances excited by the out-of-plane field component. The normalized *z*-component of the scattered electric field ($E_z^{scat}$) distributions of modes $V_1$ and $V_2$ (Fig. 2**E**) show that the modes have three field maxima in the gap but with different field profiles. Mode $V_1$ has two dominant field maxima close to the edges of the rod while, for mode $V_2$, the dominant field maximum is in the middle. These modes are the results of the anti-bonding (high energy) and bonding (low energy) hybridization, respectively, of the film-coupled vertical dipolar mode of the nanorod and the 3$^{rd}$-order (3 antinodes) Fabry-Perot mode of a metal-insulator-metal (MIM) surface plasmon polariton (SPP) supported by the nanorod-film gap[43] (Supporting Information Section 5). As the wavelength of the MIM mode is much smaller than that in the free space (approximately 60 nm for a vacuum wavelength of 650 nm), high-order MIM resonances are supported by the cavity even if it has a deep-subwavelength size. On the other hand, modes $M_2$ and $M_4$ can only be excited by $TM_1$ and $TE_2$ illuminations. The field distribution of mode $M_4$ (Fig. 2**F**) shows that it has two field maxima with an opposite charge distribution in the gap, which is due to the coupling of the longitudinal nanorod mode with the anti-phase image dipole induced in the GMF. Due to its symmetry and spectral position it is hybridized with the 2$^{nd}$-order Fabry-Perot mode of the gap. For mode $M_2$, the $E_z^{scat}$ distribution (Fig. 2**F**) indicates that it is originated from the coupling of quadrupole mode of the nanorod with the 6$^{th}$-order Fabry-Perot mode of the gap, which one can see more clearly from its $E_x^{scat}$ distribution (Supporting Information Section 6). Under $TE_1$ and $TM_2$ excitations (Fig. 2**D**), two relatively weak resonance peaks appear around 525 and 595 nm ($M_1$ and $M_3$). Together with their $E_z^{scat}$ distributions (Fig. 2**F**), one can further infer that these are film-coupled high-order transversal nanorod modes[44]. Thus, it is clear that the measured scattering peak around 550 nm is a mixture of $M_1$, $M_2$ and $M_3$ modes scattering predominantly in the top direction (Fig. 2**G**), while the peaks around 651 and 709 nm are corresponding to the excitation of vertically oriented modes of $V_1$ and $V_2$ scattering predominantly in the side directions (Fig. 2**G**).

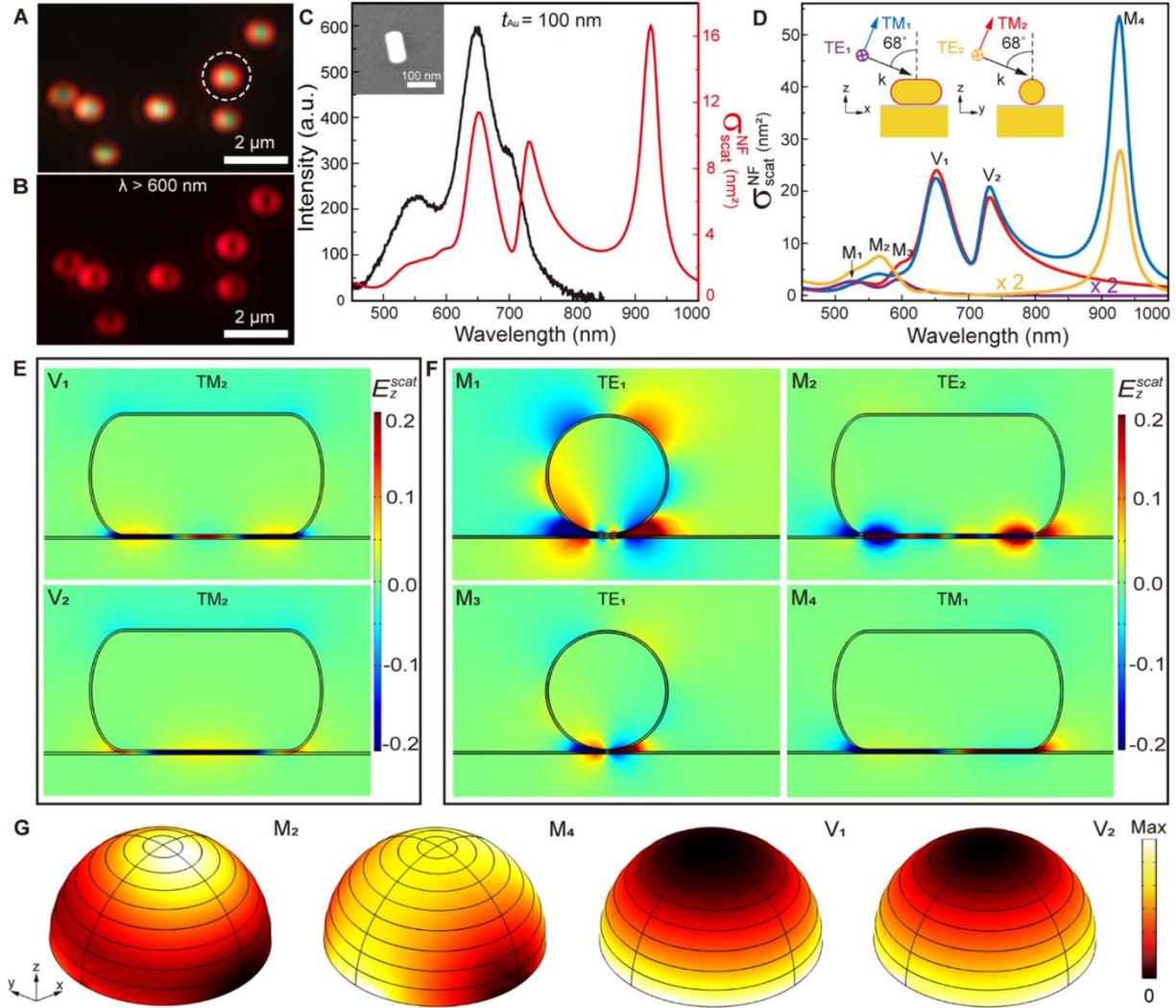

**Figure 2.** Cavity mode identification from scattering. **A,B,** Dark-field scattering images of NRoMF nanocavities formed on a GMF with $t_{Au}$ = 100 nm taken (**A**) without and (**B**) with a 600-nm long-pass filter. **C,** Measured scattering spectrum (black line) of a nanocavity in **A** (circled) and calculated partial near-field scattering cross-section $\sigma_{scat}^{NF}$ (red line) into a 50° collection angle, corresponding to the NA of the experimentally used objective. Inset: corresponding SEM image of the nanocavity. **D,** Calculated $\sigma_{scat}^{NF}$ of the NRoMF nanocavity under four different excitation conditions shown in the inset (note that the TE curves were magnified 2 times). **E,F,** Normalized $E_z^{scat}$ distributions corresponding to cavity modes labeled in **D**. **G,** Calculated energy angular scattering patterns of the nanocavity with the nanorod axis in the *x* direction, at the resonances corresponding to modes $M_2$, $M_4$, $V_1$ and $V_2$, respectively, at the same illumination condition as in **E** and **F**.

In order to compare with the experimental result (Fig. 2**C**, black line) measured under oblique illumination with an unpolarized white light (Fig. 1**A**), a partial scattering spectrum of a nanocavity (Fig. 2**C**, red line) was obtained by averaging scattering spectra calculated under four different excitation conditions (Fig. 2**D**). The calculated spectrum reproduces the experimental features very well, except the presence of an additional peak around 928 nm, which is due to the limited measurement capability of the setup at wavelength larger than 850 nm. This peak is experimentally observed for structures with an increased gap thickness (Supporting Information Section 7).

**Mirror-thickness dependent optical response.** Usually, NPoM nanocavities are constructed on metal films with thickness larger than the light penetration depth of a metal mirror[12-19,21-23,25,26] (mainly due to the significantly deteriorated surface quality of thin metal films). The ability to fabricate high-quality optically-thin (down to ~10 nm) GMFs allows the investigation of mirror-thickness dependent optical response of nanocavities. The thickness dependence should be important especially for the hybrid modes, affecting both coupling of the nanorod modes with the film and MIM Fabry-Perot modes of the gap when the mirror thickness becomes smaller than the skin depth: for thicker films MIM SPP modes are involved, while for thinner ones, insulator-metal-insulator (IMI) SPPs of the thin film become important. Figures 3**A**-**C** show optical microscope and AFM images of GMFs with thicknesses of 53, 25 and 15 nm, respectively. The transmittance of GMFs increases gradually from 25% to 48% and 70% (around 510 nm, Supporting Information Section 8) with the decrease of GMF thickness. When the GMF thickness is 53 nm, the scattering image (Fig. 3**D**,**E**) and spectrum (Fig. 3**J**, blue line) of the NRoMF nanocavities are quite similar to those formed on GMFs with 100-nm thickness (Fig. 3**J**, yellow line) except a slight decrease in the intensity. However, with the further decrease of GMF thickness down to 25 nm (Fig. 3**F**,**G**) and 15 nm (Fig. 3**H**,**I**), the doughnut-shaped scattering component fades out quickly while the inner green scattering component keeps almost unchanged. Accordingly, the intensities of resonance peaks around 660 and 710 nm decrease quickly with the decrease of GMF thickness (Fig. 3**J**), while the intensity of the resonance peak around 560 nm keeps almost constant.

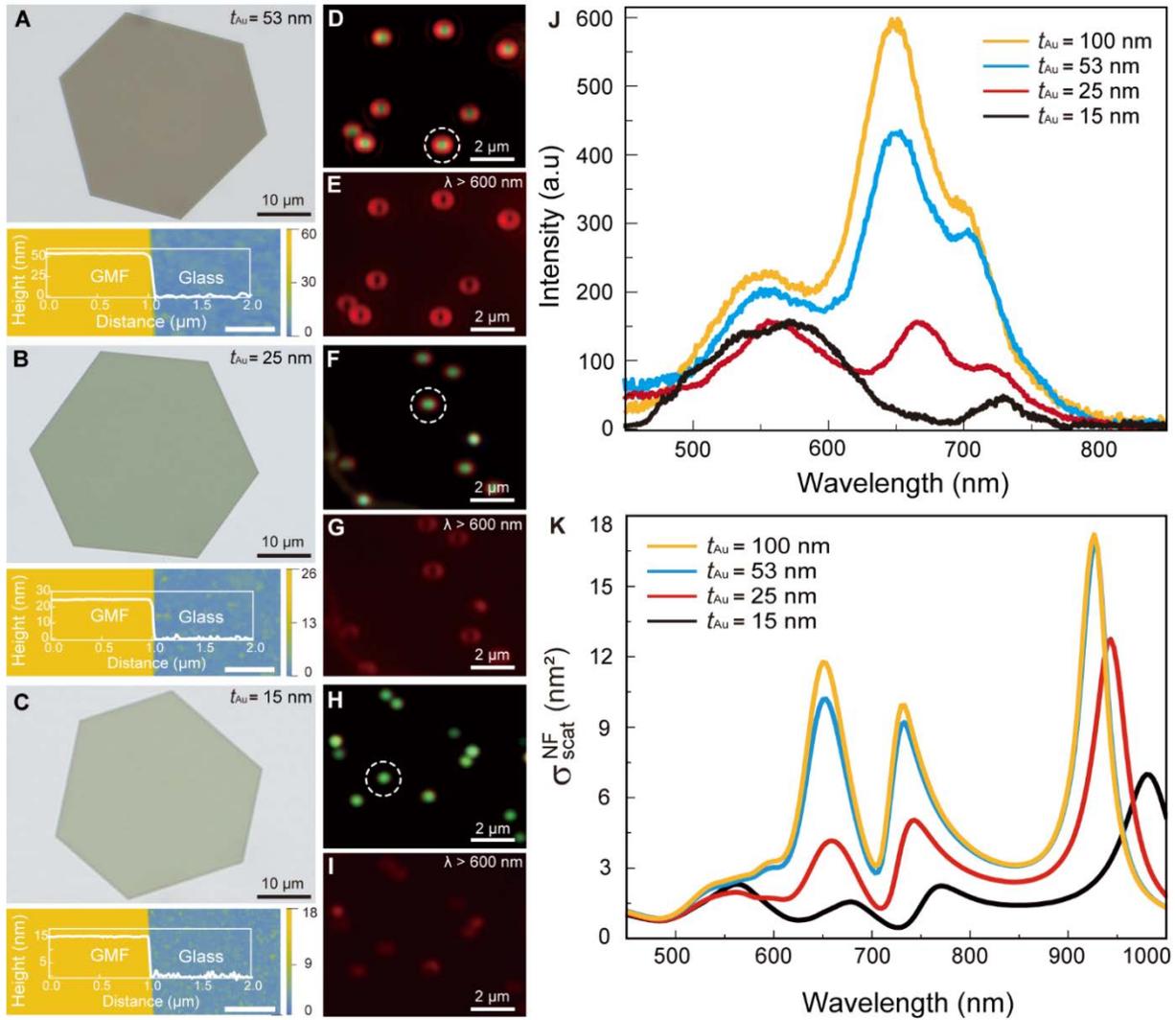

**Figure 3. Effect of GMF thickness. A-C,** Optical micrographs of GMFs with $t_{Au}$ of (**A**) 53, (**B**) 25 and (**C**) 15 nm, taken in transmission. Insets: corresponding AFM images and height profiles. Scale bar, 500 nm. **D-I,** Dark-field scattering images of the NRoMF nanocavities formed on the GMFs in **A-C** taken (**D,F,H**) without and (**E,G,I**) with a 600-nm long-pass filter. **J,** Scattering spectra of single nanocavities (circled) shown in **D,F,H** and Fig. 2A. **K,** Calculated $\sigma_{scat}^{NF}$ for the nanocavities with the GMF thickness of 100, 53, 25 and 15 nm.

The calculated evolution of near-field scattering spectra of NRoMF nanocavities with decreasing GMF thicknesses (Fig. 3**K**) agree well with the experimental observation. With the decrease of the GMF thickness, the scattering intensities of modes $V_1$ and $V_2$ decrease first slowly (from 100 to 53 nm) and then dramatically down to ~1/8 of the original values, while the intensity of the scattering peak around 560 nm stays almost at the same level. It can also be noted that as the GMF thickness gets thinner, the scattering intensity of mode $M_4$ decreases gradually (~1/3 of the original value), along with a broadening in the linewidth. The significant modification of far-field scattering properties of nanocavities with the decrease of the thickness of optically-thin GMFs can be understood as follows. Mode $M_2$ induces an antisymmetric charge distribution across the flake in the vicinity of the nanorod (see Fig. S11 and Fig. S12**A**), which with the opposite charges at the top and the bottom sides of the flake does not screen the fields of the quadrupolar component of the mode, so the latter effectively takes the form of a longitudinal dipolar mode of the nanorod radiating well both towards the collection objective and into the substrate with the decrease of GMF thickness. At the same time, modes $V_1$, $V_2$ and $M_4$ induce a symmetric charge distribution across the flake (Fig. S11) which on one hand keeps the nature of the mode the same and on the other hand couples well to the short-range IMI SPP mode of the flake (Fig. S12**B**-**D** for 15 nm thickness) as well as to the radiation into the substrate (see also Fig. S11), thus resulting in the decrease of the radiation into the objective side. Therefore, this represents a new mechanism for engineering optical properties of NPoM nanocavities. Furthermore, optically-thin GMFs provide an opportunity for the excitation of nanocavities from the mirror side via total internal reflection approach with high compactness or the out-coupling of nanocavity-enhanced optical emission from the mirror side for integration, due to the increased penetration of modes into the substrate (Fig. S11).

**Effect of mirror quality.** Due to the extreme confinement of optical fields in the nanometer-scale gap,

NPoM nanocavities are very sensitive to the structure quality of the bottom metal mirrors[34]. We compared the performance of the nanocavities obtained with granular polycrystalline gold films (Fig. 4**A**, left panel) and ultrasmooth single-crystalline GMFs (Fig. 4**B**, left panel) with thicknesses of 15, 25 and 53 nm, respectively. The superior surface quality of GMFs over the thermally deposited gold films (Supporting Information Section 10) can be clearly seen by comparing their SEM and AFM images. The former is smooth and continuous while the latter is rough and discontinuous for small thicknesses (Fig. S13). The measured RMS of the 15-nm-thickness GMF (~0.2 nm, Fig. 4**B**) is much lower than that of the deposited gold films with thicknesses of 15, 25, 53 nm (~4.1, 2.7 and 1.4 nm, respectively, Fig. 4**A** and Fig. S14). As expected, there is a clear reduction in the linewidth of scattering peaks (Figs. 4**C**-4**E**). More specifically, the quality factors of modes $V_1$ and $V_2$ for GMF-based nanocavities are about twice the values of deposited film-based nanocavities when the mirror thicknesses are 25 and 15 nm (labeled in Fig. 4**D,E**). At the same time, with the decrease of film thickness from 53 nm to 25 and 15 nm, the ratio of scattering intensities between the two cases (GMF vs. deposited film) increases from ~1.3 to 2.8 and 3.2, respectively. The great improvement in the optical qualities of NRoMF nanocavities by the use of ultrasmooth single-crystalline GMFs can be attributed to the reduction in electron scattering losses introduced by surface roughness and grain boundaries which are significant in granular polycrystalline deposited gold films. In addition, compared with nanocavities formed on rough deposited film, nanocavities formed on GMFs have a much more uniform dielectric gap thickness across the whole cavity, which is also beneficial for achieving optical modes closer to the designed values and for the decrease of an inhomogeneous broadening linewidth in a nanoparticle assemble. Thus, optically-thin GMFs with single-crystalline structure and excellent surface smoothness provide an idea platform for the construction of low-loss nanocavities that approach the theoretical limit (Supporting Information Section 11).

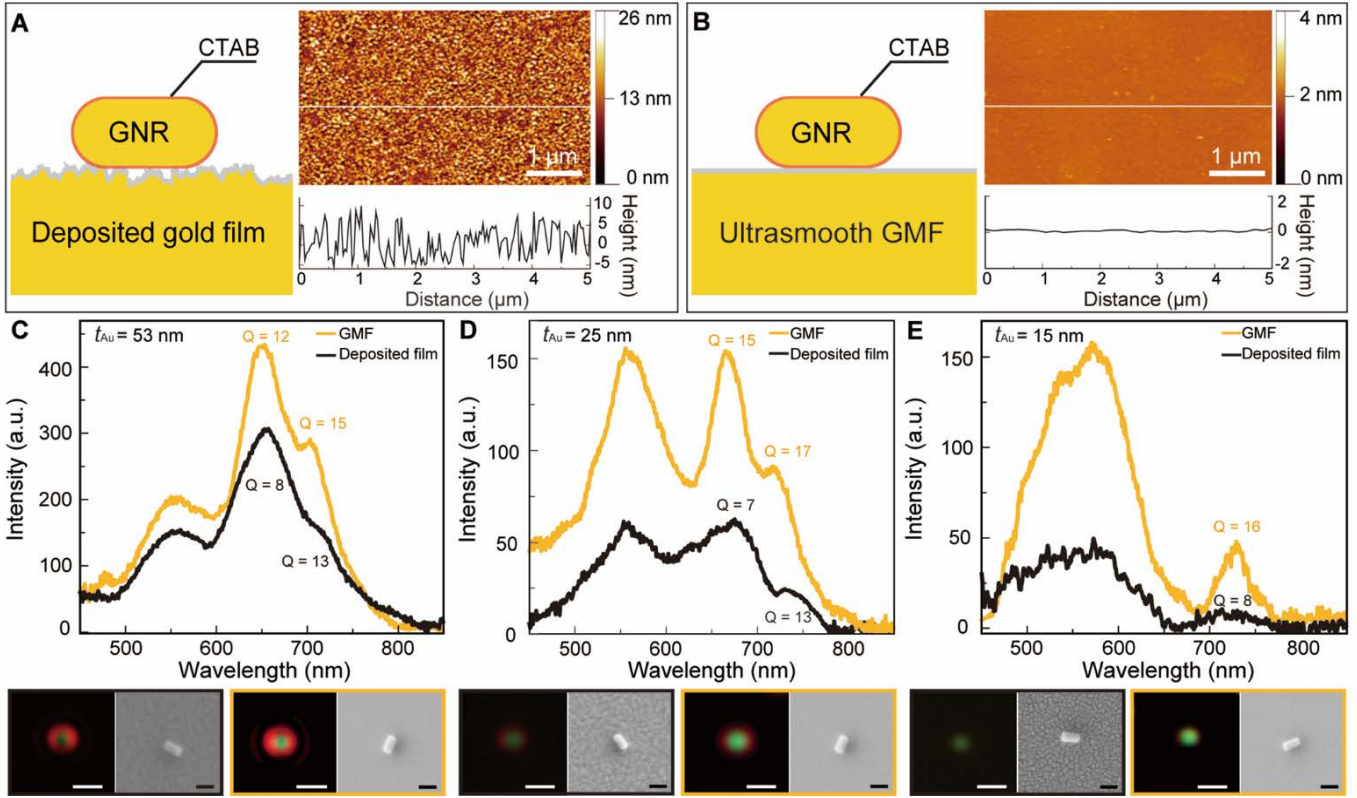

**Figure 4. Effect of mirror quality. A,B,** Schematic illustrations of nanocavities formed on (**A**) a deposited gold film and (**B**) an ultrasmooth GMF. AFM images and line scans along the indicated white lines for a deposited gold film and a GMF with 15-nm thickness are also shown. **C-E,** Comparison of the scattering spectra of the nanocavities formed on gold films and GMFs with $t_{Au}$ of (**C**) 53 nm, (**D**) 25 nm and (**E**) 15 nm. The quality factors of modes $V_1$ and $V_2$ are labeled correspondingly. Insets: the corresponding dark-field scattering and SEM images for the measured nanocavities on gold films (left panels) and GMFs (right panels), respectively. Scale bars in dark-field scattering images, 1 μm; Scale bars in SEM images, 100 nm.

**Integration with optical fibers.** To date, NPoM nanocavities are commonly exited with a bulky dark-field scattering setup[4], making it difficult to realize miniaturized nanocavity-based photonic devices. Benefiting from the transferable property of optically-thin and double-sided ultrasmooth GMFs, low-loss NRoMF nanocavities could be readily integrated with the mature platform of fiber optics and excited via total internal reflection approach with high compactness. Figure 5**A** show a schematic illustration of the integration of nanocavities with a cleaved optical fiber. Experimentally, we used a 37°-polished angled fiber (Fig. 5**B**) with a good surface quality (inset) fabricated from a standard optical fiber (Corning SMF-28e). To construct nanocavites on the surface, a GMF with a thickness of 40 nm was first transferred onto the fiber core (Fig. 5**C** and Supporting Information Section 12), followed by drop-casting of a diluted GNR solution to obtain sparsely dispersed nanocavities. Upon the launching of an unpolarized white light into the optical fiber, clear scattering spots were observed from the surface of the GMF (Fig. 5**D,E**), which have the similar scattering pattern and spectrum (Fig. 5**F**) with nanocavities excited by a dark-field spectroscopy (Fig. 3). Nanocavities can also be integrated with a microscale waveguide, e.g., a silica microfiber (Fig. 5**G**). Figure 5**H** presents an optical micrograph of a silica microfiber (20 μm in diameter) covered with a GMF (~20 nm in thickness) conformally on the sidewall (see Supporting Information Section 13 for details). Sparsely dispersed nanocavities can be efficiently excited by the waveguided unpolarized white light in the microfiber, as indicated by the scattering spots on the GMF (Fig. 5**I**). Although fiber-integrated nanocavities can also be realized by the use of deposited gold films, the approach demonstrated here provides straightforward advantages including superior optical quality, easier fabrication and lower cost.

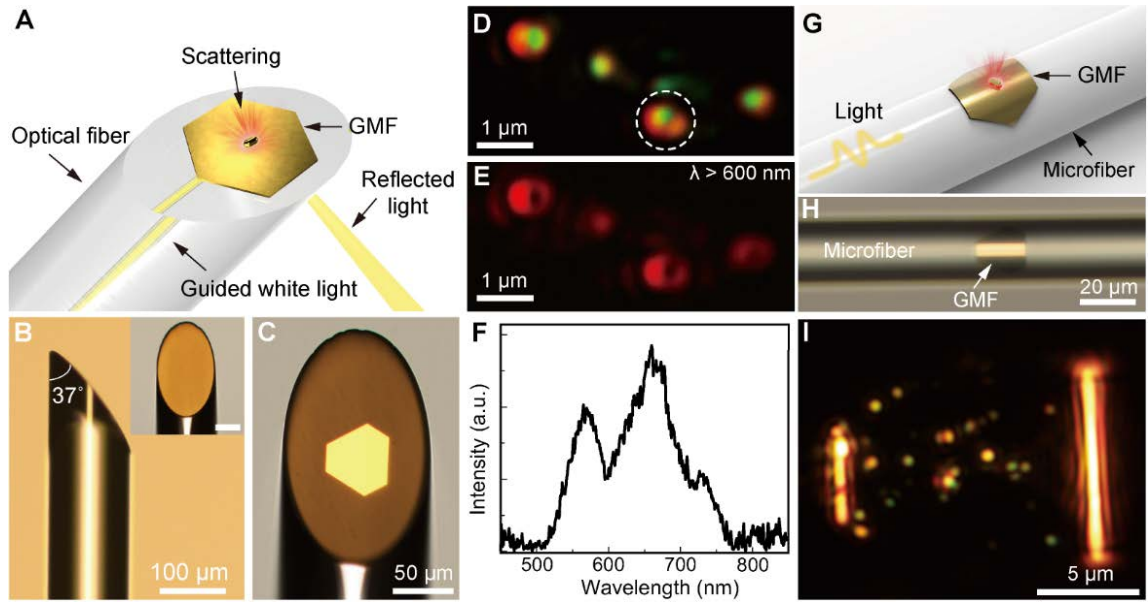

**Figure 5.** Integration with optical fibers. **A**, Schematic diagram of the integration of NRoMF nanocavities with an angled fiber. **B**,**C**, Optical micrographs of (**B**) a 37°-polished angled fiber, its polished surface (inset, the scale bar is 50 μm) and (**C**) the polished surface after transferring a GMF onto the fiber core. **D**,**E**, Scattering images of fiber-excited NRoMF nanocavities taken (**D**) without and (**E**) with a 600-nm long-pass filter. **F**, Scattering spectrum of a nanocavity (circled) in **D**. **G**, Schematic illustration of the integration of NRoMF nanocavities with a silica microfiber. **H**, Optical micrograph of a GMF on the sidewall of a microfiber. **I**, Scattering image of microfiber-integrated NRoMF nanocavities.

## CONCLUSIONS

We have shown that single-crystalline GMFs with atomically smooth surface and tunable thickness offer an ideal platform for integratable low-loss NPoM nanocavities. The superior optical quality and mode tunability of nanocavities afforded by atomically smooth single-crystalline GMFs may open new opportunities for pushing the limits of plasmonic nanocavity-based techniques and inspiring plasmonic devices with high performance. Also, the ability to integrate extreme nanophotonic platform of NPoM nanocavites with the mature platform of fiber optics or micro-/nanowaveguides, with excitation through the optically-thin ultrasmooth mirror, makes it possible to develop miniaturized nanocavity-based photonic devices, such as sensors, modulators and light sources, with high performance and low cost.

## AUTHOR INFORMATION

**Corresponding Authors**

* Email: nanopan@zju.edu.cn, a.zayats@kcl.ac.uk, phytong@zju.edu.cn.

**Author Contributions**

P.W. and L.M.T. conceived the study and co-supervised the project with A.V.Z. L.F.L. constructed the experiment and performed the measurements. C.X.Y.P synthesized gold microflakes under supervision of X.F.W. and B.H., and Y.B.T. transferred gold microflakes onto optical fibers. J.L.L. deposited the gold films. A.V.K. and J.S.Z. performed numerical simulations. P.W., A.V.K., A.V.Z. and L.M.T. analyzed the data. All the authors discussed the results and co-wrote the paper. All authors have given approval to the final version of the manuscript.

**Funding Sources**

This research was supported by the National Natural Science Foundation of China (62075195 and 12004333), National Key Research and Development Project of China (2018YFB2200404), ERC iCOMM project (789340) and Fundamental Research Funds for the Central Universities.

**Notes**

The authors declare no competing financial interest

# Supporting Information

# Atomically smooth single-crystalline platform for low-loss plasmonic nanocavities

Lufang Liu[1,#], Alexey V. Krasavin[2,#], Junsheng Zheng[1], Yuanbiao Tong[1], Pan Wang[1,*], Xiaofei Wu[3], Bert Hecht[3], Chenxinyu Pan[1], Jialin Li[1], Linjun Li[1], Xin Guo[1], Anatoly Zayats[2,*], Limin Tong[1,*]

[1]State Key Laboratory of Modern Optical Instrumentation, College of Optical Science and Engineering, Zhejiang University, Hangzhou 310027, China

[2]Department of Physics and London Centre for Nanotechnology, King's College London, Strand, London WC2R 2LS, UK

[3]NanoOptics & Biophotonics Group, Experimentelle Physik 5, Physikalisches Institut, Universität Würzburg, Am Hubland, 97074 Würzburg, Germany

[#]These authors contributed equally to this work

***Corresponding Authors**. Email: nanopan@zju.edu.cn, a.zayats@kcl.ac.uk, phytong@zju.edu.cn

**Section 1. Synthesis of GMFs.**

GMFs were synthesized using a modified wet-chemical method[1]. First, a growth solution was prepared by the addition of 90-μL of 0.1 M chloroauric acid (98%, Sigma-Aldrich) aqueous solution into a 10-mL ethylene glycol (99.8%, Sigma-Aldrich) in a 20-mL glass vial. Then, a cleaned glass slide was immersed into the solution at a slightly tilted angle. Finally, the growth solution was heated to 95°C in an oven and kept at this temperature for a certain reaction time defining the dimensions of the resulting GMFs. After the growth, the glass slide with GMFs on the surface was taken out of the growth solution, cleaned with ethanol, and dried in a nitrogen environment for the next-step use. The thickness of GMFs can be tuned from ~10 nm to hundreds of nanometers by the control of growth time from ~4 to 24 h. As shown in Fig. S1, with the increase of growth time, the size of GMFs increases from ~10 μm to hundreds of micrometers. At the same time, there is a gradual change in the colour of GMFs taken both in reflection and transmission, which is due to the increase of the GMF thickness.

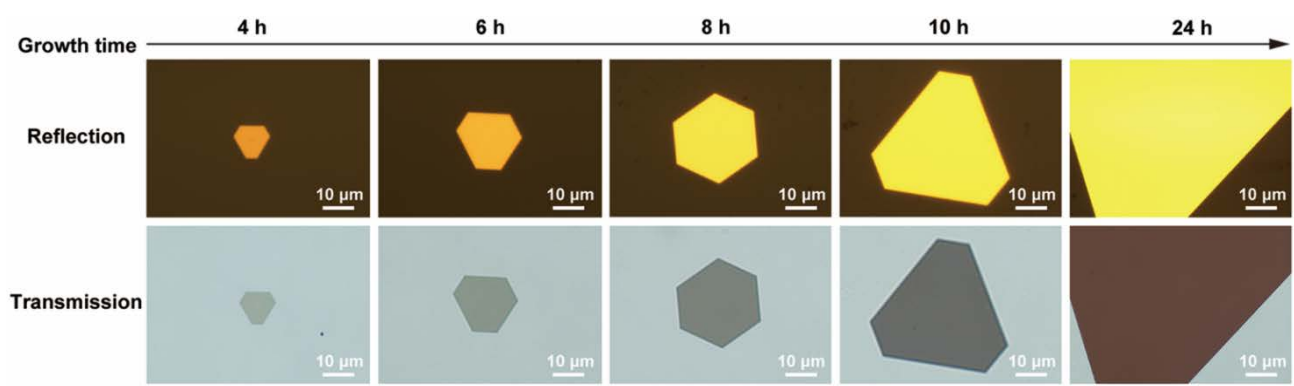

**Figure S1.** Optical micrographs of GMFs with an increasing growth time from 4 to 24 h. The top panels were taken in reflection, while the bottom panels were taken in transmission.

## Section 2. Synthesis of GNRs

GNRs were prepared using a seed-mediated method[2]. Firstly, to prepare the seed solution, an ice-cold NaBH$_4$ solution (0.6 mL, 0.01 M) was added into a mixture of HAuCl$_4$ (0.25 mL, 0.01 M) and CTAB (9.75 mL, 0.1 M) aqueous solutions. The resultant solution was rapidly stirred for 2 min and kept at room temperature for 2 h before use. Secondly, the growth solution was made by the sequential addition of HAuCl$_4$ (2 mL, 0.01 M), AgNO$_3$ (0.4 mL, 0.01 M), ascorbic acid (0.32 mL, 0.1 M) and HCl (0.8 mL, 1.0 M) aqueous solutions into a CTAB aqueous solution (40 mL, 0.1 M). The obtained solution was mixed by gently shaking. Then 3 μL seed solution was added to the growth solution. The resultant solution was gently stirred for 20 s and then kept undisturbed overnight. Figure S2**A** shows a TEM image of as-synthesized GNRs. The average diameter and length are determined to be 55 ± 5 and 105 ± 6 nm, respectively. Figure S2**B** shows an extinction spectrum of GNRs dispersed in water.

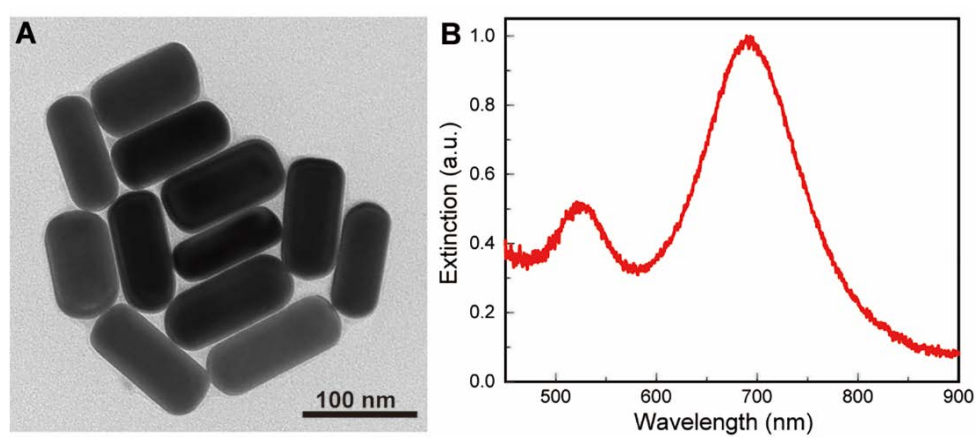

**Figure S2.** TEM images and extinction spectra. **A**, TEM image of GNRs. **B**, Extinction spectrum of GNRs dispersed in an aqueous solution.

## Section 3. Dark-field spectroscopy setup

Scattering of single nanocavities was characterized via a dark-field scattering spectroscopy, as schematically shown in Fig. S3. Briefly, an unpolarized white light from a halogen tungsten lamp was first focused onto NRoMF nanocavities on a glass slide at an incident angle of 68° by a 100× dark-field objective (NA = 0.8, TU Plan ELWD, Nikon). The scattering light from single NRoMF nanocavities was collected by the same objective and directed with a beam splitter to a charge-coupled device camera (DS-Fi3, Nikon) for imaging and a spectrometer (QE pro, Ocean Insight) for the spectral analysis. All the scattering spectra were measured under the same illuminating condition and integration time.

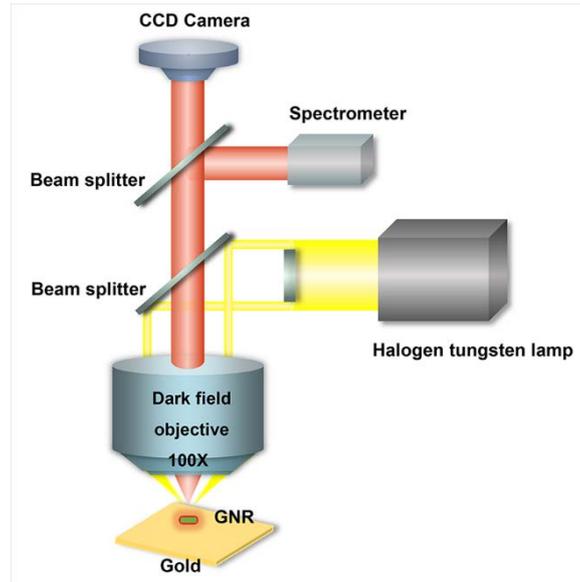

**Figure S3**. Schematic diagram of the dark-field spectroscopy setup.

## Section 4. Numerical simulations

The simulations were performed using a finite element method (COMSOL Multiphysics 5.4) in a scattered field formulation. GNRs implemented as elliptically-tapered cylinders matching the geometry of their experimental counterparts and covered with a CTAB layer (1 nm in thickness) were placed with proper spacers onto a gold slab presenting the microflake lying on a substrate. The nanostructure was illuminated with a plane wave at 68°. To avoid back-reflection, the simulation domain was surrounded by perfect matched layer. To reduce the numerical complexity of the simulations symmetry of the problem was utilized, which allows modeling of only a half of the overall domain setting appropriate (perfect electric or perfect magnetic conductor) boundary conditions on the slicing boundary, defined by the polarization of the incident wave. The wavelength of the incident wave was varied from 450 to 1000 nm, while the power flow of the scattered fields was integrated in the near-field region 350 nm from the nanorod center and inside a 50° collection angle corresponding to the NA of the objective employed in the experiments. The partial near-field scattering cross-section $\sigma_{\text{scat}}^{\text{NF}}$ was obtained by dividing the obtained integral with the intensity of the incident wave. The refractive indices of gold, CTAB were taken from ref. 3 and 4 , respectively. While the refractive index of single-crystalline optically-thin flakes may deviate from the tabulated values, their thickness dependence will be investigated elsewhere.

## Section 5. Formation of modes $V_1$ and $V_2$

For a nanosphere-on-mirror nanocavity under TM excitation, a vertically (perpendicular to a mirror) polarized mode (Fig. S4) can be excited due to the coupling of the vertical dipolar mode of the nanosphere with its image dipole in the gold mirror[5], which has only one hotspot at the center of the gap. When the nanosphere is elongated into a rod shape, in addition to the film-coupled vertical dipolar mode, the nanocavity can also support in this wavelength range nonradiative 3$^{rd}$-order (3 antinodes) Fabry-Perot resonance in the MIM gap defined by the length of the rod[6]. Thus, the hybridization between the film-coupled vertical dipolar mode and the 3$^{rd}$-order Fabry-Perot mode forms antibonding and bonding modes (Fig. S5), named modes $V_1$ and $V_2$, respectively.

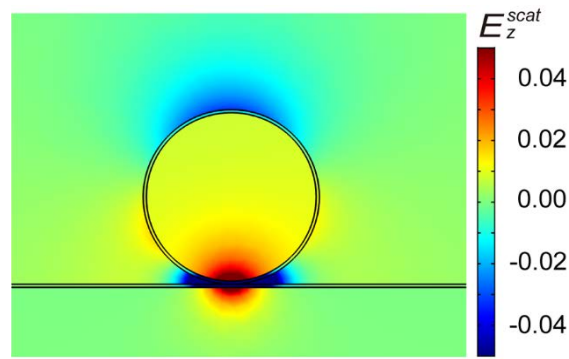

**Figure S4.** Normalized $E_z^{scat}$ field distribution of film-coupled vertical dipolar mode of a nanocavity formed by a 55-nm-diameter gold nanosphere on a GMF (100 nm in thickness).

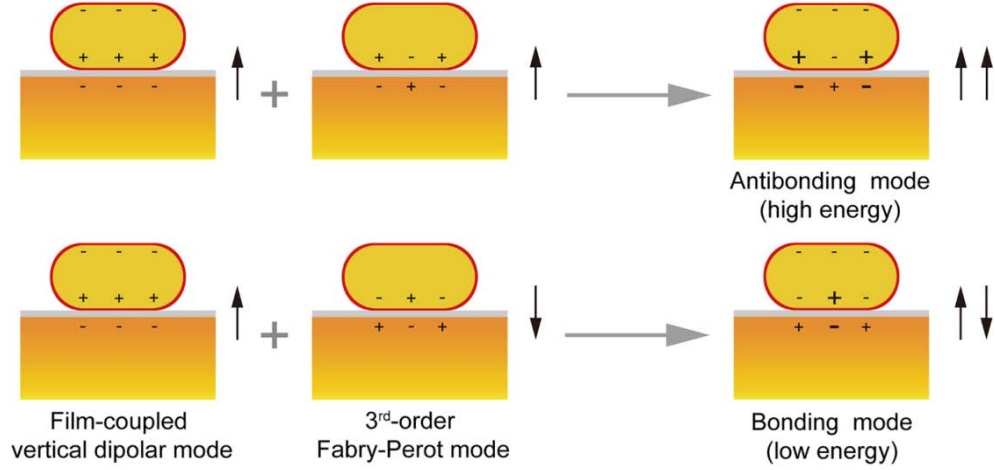

**Figure S5.** Schematic diagram showing hybridization between film-coupled vertical dipolar mode and 3rd-order Fabry-Perot mode to generate modes $V_1$ and $V_2$.

## Section 6. $E_x^{scat}$ field distribution of mode $M_2$

Figure S6 presents the $E_x^{scat}$ field distribution of mode $M_2$ of the NRoMF nanocavity investigated in Fig. 2. Note that the charge maxima of the Fabry-Perot mode correspond to the node of the $E_x^{scat}$ field, so the Fabry-Perot mode in the gap is clearly of the 6th order.

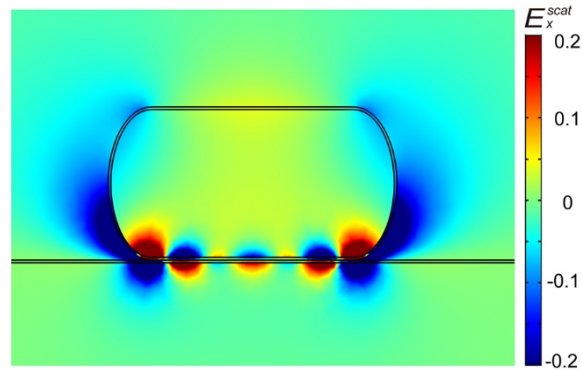

**Figure S6.** Normalized $E_x^{scat}$ field distribution of mode $M_2$ of the nanocavity investigated in Fig. 2.

## Section 7. Effect of spacer thickness on the optical properties

The effect of spacer thickness on the optical properties of NRoMF nanocavities was investigated by employing an amorphous $Al_2O_3$ layer as the dielectric spacer. $Al_2O_3$ thin layers with various thicknesses were deposited on 100-nm-thickness GMFs using atomic layer deposition (ALD, SENTECH SI ALD) at 120 °C. To initiate the first reaction cycle on the GMFs, the surface of the GMFs were hydroxylated by immersing GMF-grown glass slides into a 10 mM 2-mercapto-ethanol (99%, Sigma-Aldrich) ethanol solution overnight. Then the glass slides were rinsed with excessive ethanol and dried with nitrogen for ALD deposition. The thickness of $Al_2O_3$ spacer layer can be precisely controlled by controlling the reaction cycle. Figure S7 presents TEM images of the edges of $Al_2O_3$ coated GMFs, clearly showing the thickness of $Al_2O_3$ layer to be 1.5, 2.4 and 4.0 nm.

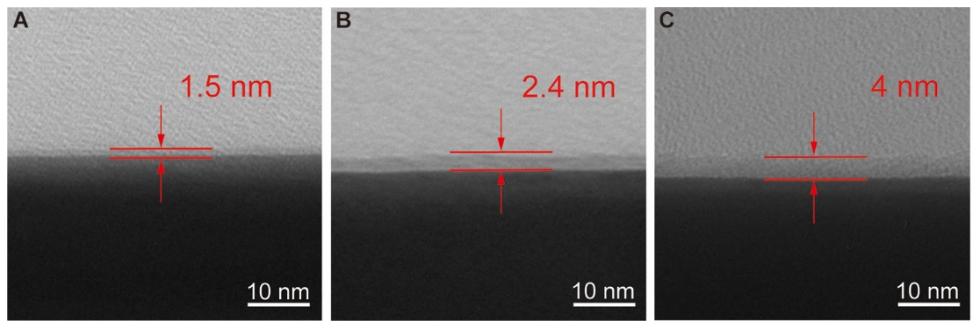

**Figure S7**. TEM images of the edges of GMFs with different thicknesses of $Al_2O_3$ layers: (**A**) 1.5, (**B**) 2.4 and (**C**) 4 nm.

When the $Al_2O_3$ thickness is 1.5 nm, as have been observed in the main text, there are three resonance peaks in the scattering spectrum (Fig. S8, yellow line), located at 550, 610 and 700 nm, respectively. With the further increase of $Al_2O_3$ thickness to 2.4 and 4.0 nm (green and blue lines), along with the blue-shift of the resonance peaks of modes $V_1$ and $V_2$, a new resonance peak (around 811 and 783 nm, respectively) emerges. This peak corresponds to mode $M_4$ predicted in Fig. 2**C**, which now moves into the detection spectral range of the set-up. Simultaneously, the scattering intensity of mode $M_4$ becomes stronger as the nanorod dipole and its image in metal get more separated. Accordingly, with the increase of $Al_2O_3$ thickness, there is a distinct change in the scattering patterns of NRoMF nanocavities (insets of Fig. S8). It should be mentioned that the relatively weak scattering intensity of mode $M_4$ observed experimentally is mainly due to the chromatic aberration of the optical system[7].

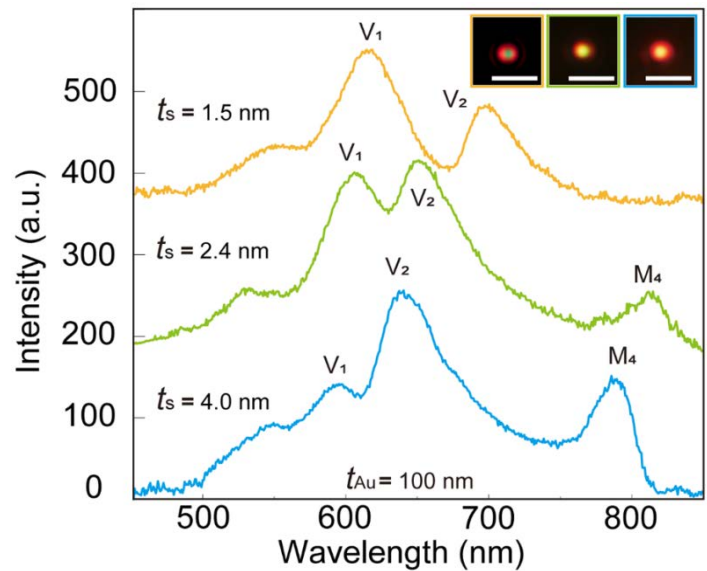

**Figure S8.** Scattering spectra of NRoMF nanocavities formed on 100-nm-thickness GMFs with varying $Al_2O_3$ thickness $t_s$ of 1.5, 2.4 and 4.0 nm. Insets: corresponding dark-field scattering images of the measured nanocavities. Scale bar, 2 μm.

## Section 8. Transmittance of GMFs with different thicknesses

The transmittance of GMFs was measured with a white light focused in the center of a GMF under an optical microscope. As shown in Fig. S9, with the decrease of a GMF thickness, the transmittance of GMFs increases gradually from 25% to 48% and 70% at the wavelength of 510 nm.

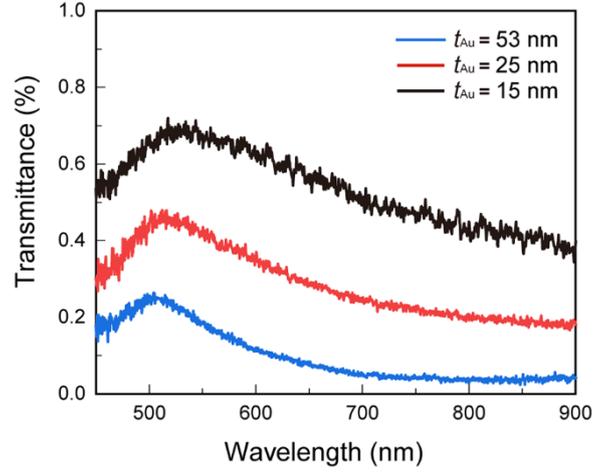

**Figure S9.** Transmittance of GMFs with thicknesses of 53, 25 and 15 nm, respectively.

# Section 9. Scattering cross-sections and field distributions of nanocavities on GMFs with various thicknesses

Figure S10 presents calculated scattering cross-sections of NRoMF nanocavities formed on GMFs with thicknesses of 100, 53, 25, 15 and 6 nm, respectively. The corresponding normalized $E_z^{scat}$ field distributions of modes $M_2$, $M_4$, $V_1$ and $V_2$ of NRoMF nanocavities are further shown in Fig. S11.

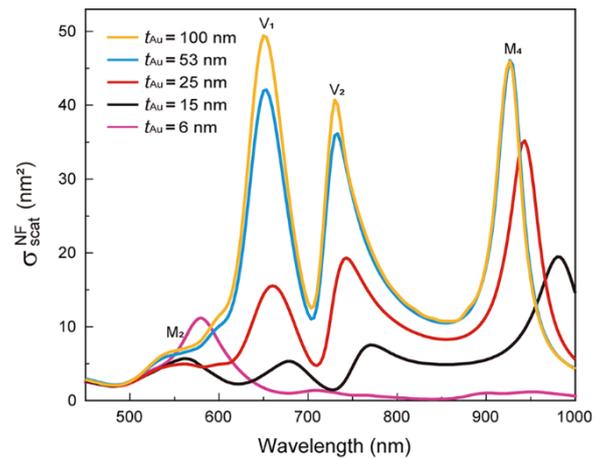

**Figure S10.** Calculated $\sigma_{scat}^{NF}$ for NRoMF nanocavities formed on GMFs with thicknesses of 100, 53, 25, 15 and 6 nm, respectively.

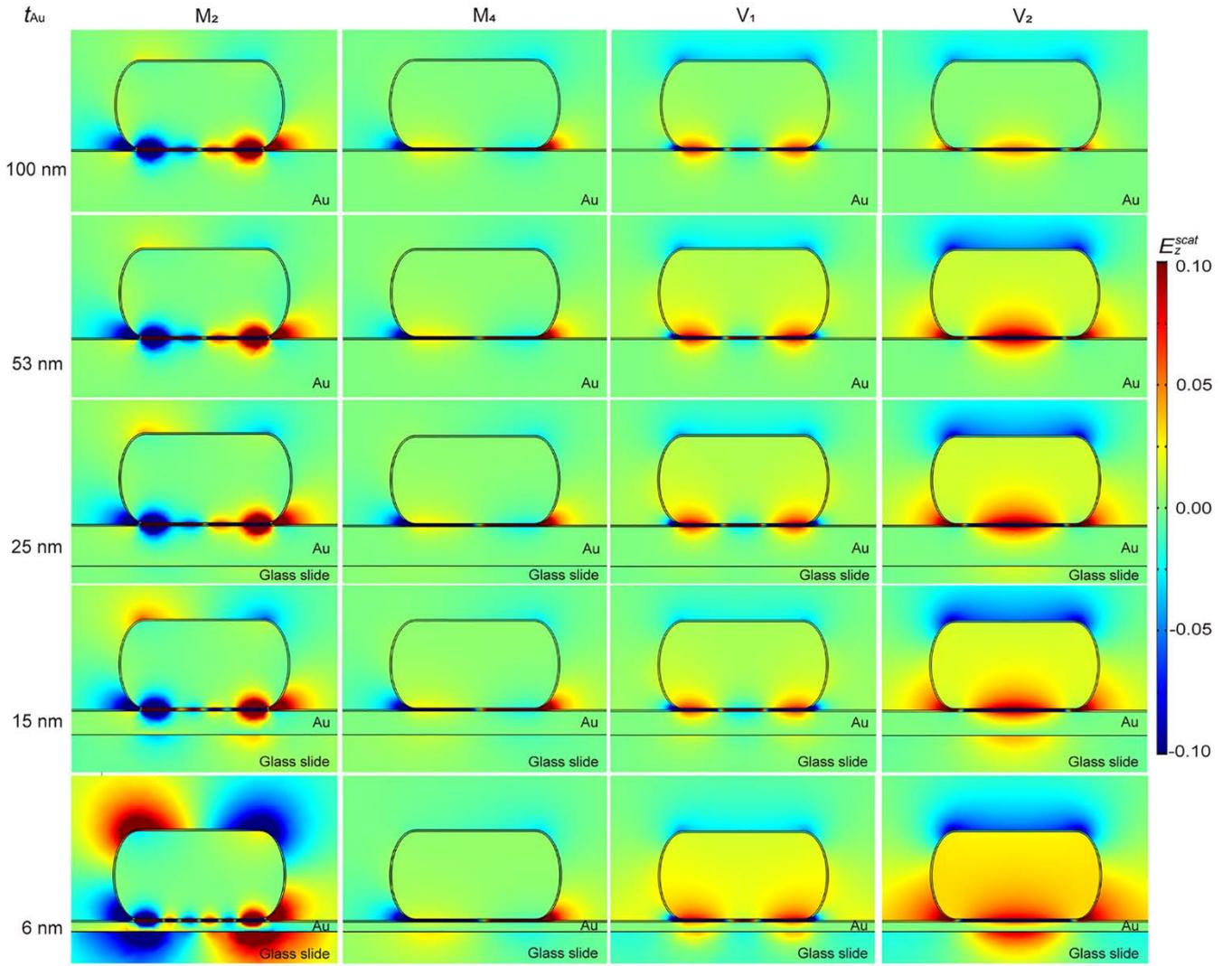

**Figure S11.** Normalized $E_z^{scat}$ field distributions of modes $M_2$, $M_4$, $V_1$ and $V_2$ for NRoMF nanocavities, as indicated in Fig. S10, with $t_{Au}$ of 100, 53, 25, 15 and 6 nm.

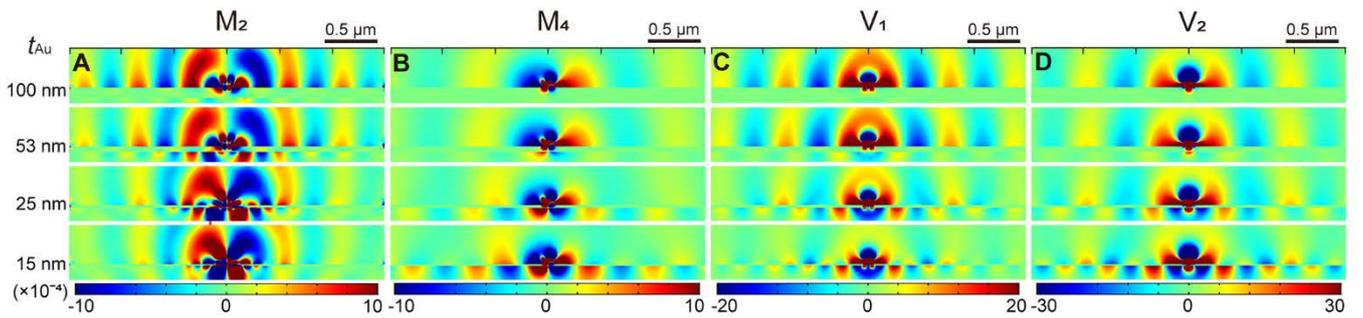

**Fig. S12.** Spatial distributions of normalized $E_z^{scat}$ field at the resonance peaks of modes $M_2$, $M_4$, $V_1$ and $V_2$ in the $x$-$z$ plane for the nanocavities formed on the GMFs of various thicknesses under $TM_1$ excitation.

Figure S12 shows the calculated spatial distributions of normalized $E_z^{scat}$ field at the resonance peaks of modes $M_2$, $M_4$, $V_1$ and $V_2$ in the *x-z* plane. As can be seen on the field distributions, in addition to couple into free space as photons, the excited plasmonic modes in NRoMF nanocavities can also couple into the GMF as propagating SPPs[8,9]. At large thicknesses (e.g., 100 nm) the NRoMF modes are coupled to the SPP mode at the gold-air interface. With the decrease of GMF thickness, SPPs propagating at the gold-air and gold-silica interfaces are coupled and form a highly-confined short-range IMI propagating mode[10,11], so the NRoMF modes may couple to it. For mode $M_2$ with a resonance at short wavelength, the short-range IMI mode is excited with low efficiency and can be observed only in the close vicinity of the nanocavity (Fig. S12**A**, bottom two panels) as in this spectral range the SPP is extremely lossy. On the other hand, for mode $M_4$ positioned at much larger wavelength, the short-range IMI SPP mode is excited by NRoMF with higher efficiency due to a better mode overlap and can propagate for a longer distance (Fig. S12**B**, bottom panel) as the loss in this spectral range is lower. Analogous situation is observed for the case of vertically oriented resonances (Fig. S12**C**,**D**).

## Section 10. Deposition and characterization of gold films

Gold films were prepared by thermal evaporation (Nano36, Kurt J. Lesker). To avoid the use of metallic adhesion layers (such as Cr, Ti) that can introduce significant optical loss to plasmonic structures[12], cleaned glass slides were first functionalized with a monolayer of (3-Aminopropyl)trimethoxysilane instead for the subsequent deposition of gold. This was realized by immersing them into a 5 mM (3-Aminopropyl)trimethoxysilane (97%, Sigma-Aldrich) ethanol solution overnight and kept at 80 °C for 1 h to enhance the attachment of (3-Aminopropyl)trimethoxysilane molecules on the glass slide surface. The functionalized glass slides were rinsed with excessive ethanol and dried with nitrogen. Gold films with different thicknesses were deposited in vacuum (~$10^{-6}$ Torr) on the prepared glass slides at a rate of 1 Å/s. Figure S13 shows SEM images of the surface of a GMF and deposited films with thicknesses of 15 and 53 nm, respectively. Figure S14 (and Figures 4**A** and 4**B**) presents AFM images of the surface of a GMF and deposited gold films. The superior surface quality of GMFs over the deposited gold films can be clearly seen.

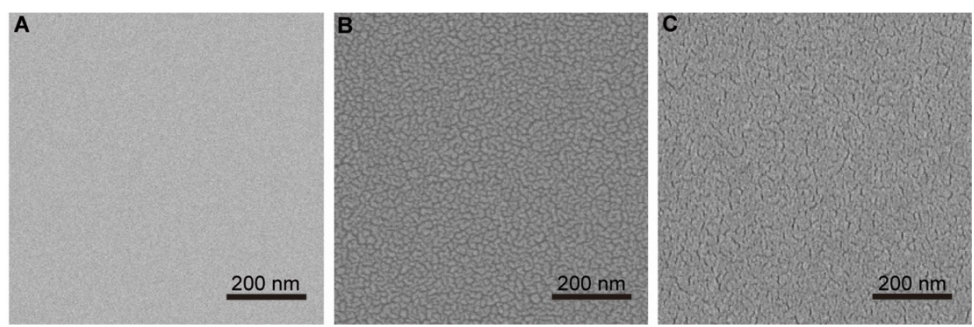

**Figure S13**. SEM images of GMF and deposited gold films. **A**, SEM image of 15-nm-thickness GMF with atomic-level surface roughness. **B**, **C**, SEM images of deposited gold films with $t_{Au}$ of (**B**) 15 and (**C**) 53 nm.

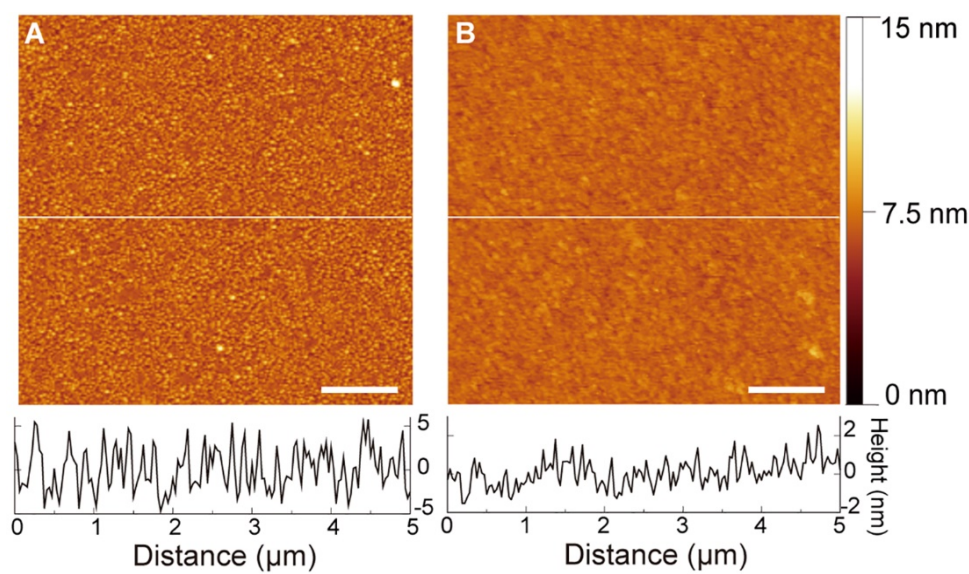

**Figure S14**. AFM images of the surface of thermally-deposited gold films with thicknesses of (**A**) 25 and (**B**) 53 nm with the RMS roughness of 2.7 and 1.4 nm, respectively.

## Section 11. Comparison of quality factors

The quality factors (Q factors) of modes $V_1$ and $V_2$ of scattering spectra for GMF-based and deposited film-based NPoM nanocavities with various mirror thicknesses (shown in Figs. 4**C**-4**E**) are shown in Table S1. The Q factors of modes $V_1$ and $V_2$ in the calculated results (shown in Fig. 3**K**) are presented as well in the table for comparison. The Q factors of modes $V_1$ and $V_2$ for GMF-based nanocavities are higher than those of deposited film-based NPoM nanocavities. When the mirror thickness is decreased to 25 or 15 nm, the Q factors of modes $V_1$ and $V_2$ for GMF-based nanocavites are about twice the values of deposited film-based nanocavities, approaching the theoretical limit.

**Table S1** Comparison of Q factors of modes $V_1$ and $V_2$ of scattering spectra for GMF and deposited film-based nanocavities as well as the calculated results.

|  | $t_{Au}$ = 53 nm | | $t_{Au}$ = 25 nm | | $t_{Au}$ = 15 nm | |
| --- | --- | --- | --- | --- | --- | --- |
|  | $V_1$ | $V_2$ | $V_1$ | $V_2$ | $V_1$ | $V_2$ |
| Deposited gold film | Q = 8 | Q = 13 | Q = 7 | Q = 13 | / | Q = 8 |
| GMF | Q = 12 | Q = 15 | Q = 15 | Q = 17 | / | Q = 16 |
| Calculated | Q = 14 | Q = 18 | Q = 14 | Q = 16 | Q = 12 | Q = 16 |

## Section 12. Transfer of GMFs onto an angled fiber

The chemically synthesized GMFs are movable, and thus can be readily transferred onto other substrates via a polydimethylsiloxane (PDMS)-mediated approach. As schematically shown in Fig. S15, a GMF grown on a glass slide was first picked up by a clean PDMS thin film, and then an angled fiber was aligned to make a contact with the GMF on the PDMS film. Finally, the PDMS film was moved away carefully, and the GMF was left on the polished surface of the angled fiber.

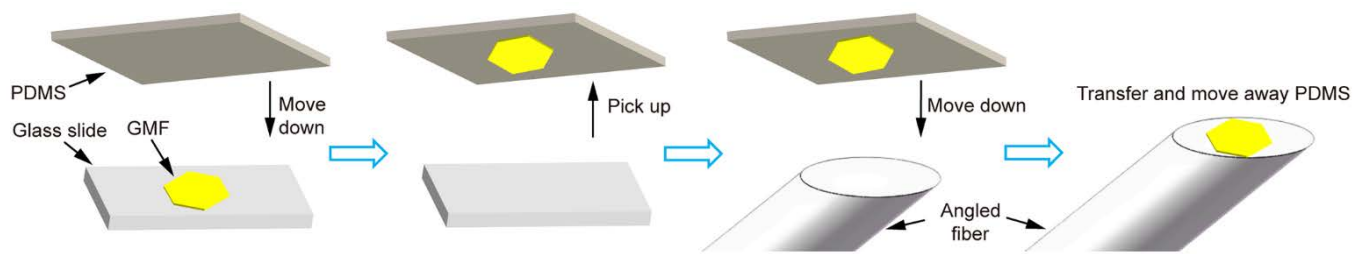

**Figure S15**. PDMS-mediated transfer of a GMF onto the polished surface of an angled fiber.

## Section 13. Integration with an optical microfiber

NRoMF nanocavities can also be readily integrated with a silica microfiber, and excited by the evanescent field of the waveguided light. Silica microfiber, with a diameter of 20 μm, was fabricated by flame-assisted taper drawing from a standard optical fiber[13]. Using the method described in Section 12, a GMF with a thickness of ~20 nm and a lateral size of 27 μm was transferred onto the sidewall of the silica microfiber. Subsequently, GNRs were deposited onto the GMF by drop-casting to obtain sparsely dispersed NRoMF nanocavities.